\documentclass[letterpaper,twocolumn,nobibnotes,aps,prb,showpacs,floatfix]{revtex4-1}\pdfoutput=1

\usepackage[utf8]{inputenc}
\usepackage[T1]{fontenc}
 
\usepackage{amsmath, amssymb, amsfonts}
\usepackage{amstext, mathrsfs, textcomp}
\usepackage{nicefrac}
\usepackage{multirow}

\usepackage{hyperref}
\usepackage{subfigure}
\usepackage{graphicx}
\usepackage{xcolor}
\graphicspath{{Figures/}}

\usepackage{notes2bib}
\bibnotesetup{
note-name
 = ,
use-sort-key = false
}

\begin{document}

\title{Origin of Second Harmonic Generation from Individual Silicon Nanowires}

\author{\firstname{Peter R.} \surname{Wiecha}$^{1,2}$}
\author{\firstname{Arnaud} \surname{Arbouet}$^{1}$}
\author{\firstname{Christian} \surname{Girard}$^{1}$}
\author{\firstname{Thierry} \surname{Baron}$^{3}$}
\author{\firstname{Vincent} \surname{Paillard}$^{1,2}$}
\email[corresponding author~: ]{vincent.paillard@cemes.fr}

\affiliation{$^{1}$ CEMES-CNRS, $Universit\acute{e}$ de Toulouse, 29 rue Jeanne Marvig, 31055 Toulouse Cedex 4, France}
\affiliation{$^{2}$ $Universit\acute{e}$ Toulouse III-Paul Sabatier, $Universit\acute{e}$ de Toulouse, 118 route de Narbonne, 31062 Toulouse cedex 9, France}
\affiliation{$^{3}$ LTM-CNRS-CEA-Universit$\acute{e}$ de Grenoble Alpes, 17 rue des martyrs, 38054 Grenoble, France}

\begin{abstract}
We investigate second harmonic generation from individual silicon nanowires and study the influence of resonant optical modes on the far-field nonlinear emission.
We find that the polarization of the second harmonic has a size-dependent behavior and explain this phenomenon by a combination of different surface and bulk nonlinear susceptibility contributions. 
We show that the second harmonic generation has an entirely different origin, depending on whether the incident illumination is polarized parallel or perpendicularly to the nanowire axis. 
The results open perspectives for further geometry-based studies on the origin of second harmonic generation in nanostructures of high refractive index centrosymmetric dielectrics.
\end{abstract}


\maketitle

\section{Introduction}
\label{int}

High-index semiconductor nanostructures attract increasing interest as low-loss alternatives to plasmonic particles.\cite{albella_low-loss_2013,albella_electric_2014, spinelli_light_2014} 
Such nanostructures provide original optical properties thanks to the occurrence of size- and shape-dependent optical resonances, which can be used to enhance and engineer light-matter interaction.\cite{brongersma_light_2014,ee_shape-dependent_2015} 
In particular scattering and absorption efficiencies or local electromagnetic field intensity can be spectrally tuned and enhanced, which opens a multitude of possible applications in photovoltaics, \cite{spinelli_light_2014} photonics,\cite{priolo_silicon_2014} or field-enhanced spectroscopies. \cite{kallel_photoluminescence_2013,albella_low-loss_2013}

In photonics, also nonlinear optical effects play an important role, offering a wide range of functionalities such as coherent up-conversion of laser light, generation of short pulses, all-optical signal processing or ultrafast switching.\cite{tanabe_fast_2007,leuthold_nonlinear_2010,aouani_third-harmonic-upconversion_2014} 
Nonlinear effects, however, are intrinsically weak. The coherent conversion of two low-energy photons into a single photon at double energy, so-called second harmonic generation (SHG), was experimentally discovered not earlier than 1961, consequently to the invention of the laser.\cite{franken_generation_1961} 
Plasmonic nanostructures, capable to strongly localize far-field radiation, are very promising candidates to boost nonlinear optical effects, hence driving the increasing interest for nonlinear plasmonics.\cite{metzger_strong_2015,hubert_role_2007,kauranen_nonlinear_2012} 
As an alternative to metal nanoparticles, enhanced nonlinear properties can be obtained in semiconductor nanostructures by taking advantage of resonant optical modes. 
For instance, intense second harmonic generation (SHG) was achieved in nanowires of III-V compounds, \cite{long_far-field_2007,grange_far-field_2012,sanatinia_modal_2014} or enhanced third harmonic generation could be produced in silicon nanostructures.\cite{jung_vitro_2009,shcherbakov_nonlinear_2015}

However, centrosymmetric materials like elemental semiconductors have a vanishing second order susceptibility, as inversion symmetry forbids even order terms in the electric polarization expansion. Second order processes can therefore occur only in the presence of interfaces or field gradients.\cite{guyot-sionnest_general_1986} 
Yet, for very small systems, the surface to volume ratio becomes high and fairly strong second order effects may arise from the broken symmetry, possibly further supported by local field enhancement due to resonant modes.

As SHG from centrosymmetric materials can be due to different processes, the source of the largest contribution to SHG has led to controversial conclusions. Often, second order effects in centrosymmetric nanostructures are modeled assuming the $\chi^{(2)}_{\perp \perp \perp}$ surface contribution from field components normal to the surface to be most significant, neglecting other possible sources.\cite{falasconi_bulk_2001,berthelot_silencing_2012}
Studies on the magnitude of other contributions have been performed on homogeneous surfaces\cite{cattaneo_polarization-based_2005,wang_surface_2009} or on nanoparticles like metal nanospheres.\cite{dadap_second-harmonic_1999,bachelier_origin_2010} 
A geometrical study on the selection rules for local surface and non-local bulk contributions to SHG from metal nano-tips under planewave excitation pointed out a purely surface-like SHG in colinear measurements.\cite{neacsu_second-harmonic_2005} 
Nonlocal bulk contributions to SHG might also be present from field gradients due to resonant modes or tightly focused beams. For instance, the influence of a field gradient has been theoretically described for low-index spherical nanoparticles excited by a tightly focused beam, and a characteristic signature in the far-field emission pattern has been predicted.\cite{huo_second_2011}

It is consequently very interesting to better understand the origin of SHG in high-index dielectric nanostructures of centrosymmetric materials, supporting a finite number of resonant modes at subwavelength dimensions.
It was recently shown that SHG from individual strain-free silicon nanowires (Si-NWs) is strongly increased if a Mie optical mode corresponding to the exciting laser wavelength is supported by the nanowire, while no SHG can be detected in absence of any resonance.\cite{wiecha_enhanced_2015} 
In fact, nonlinear scattering theory and the so-called \emph{Miller's rule} predict highest SHG for an optimum overlap of resonant modes and linear susceptibilities, at both the fundamental and harmonic wavelengths, respectively. \cite{hubert_role_2007,miller_optical_1964}
In this work, we investigate SHG from individual Si-NWs, which can be interesting for silicon-based nanophotonics, and address the different sources of SHG in such systems. 
We show that some of our experimental observations cannot be explained by $\chi^{(2)}_{\perp \perp \perp}$-SHG, and find that contributions from tangential fields at the surface as well as from strong field gradients in the bulk have to be considered, depending on the NW diameter.

\section{Silicon Nanowire Samples and Experimental Setup}
\label{exp}

\begin{figure}[tb]
\centering
\includegraphics[width=\columnwidth]{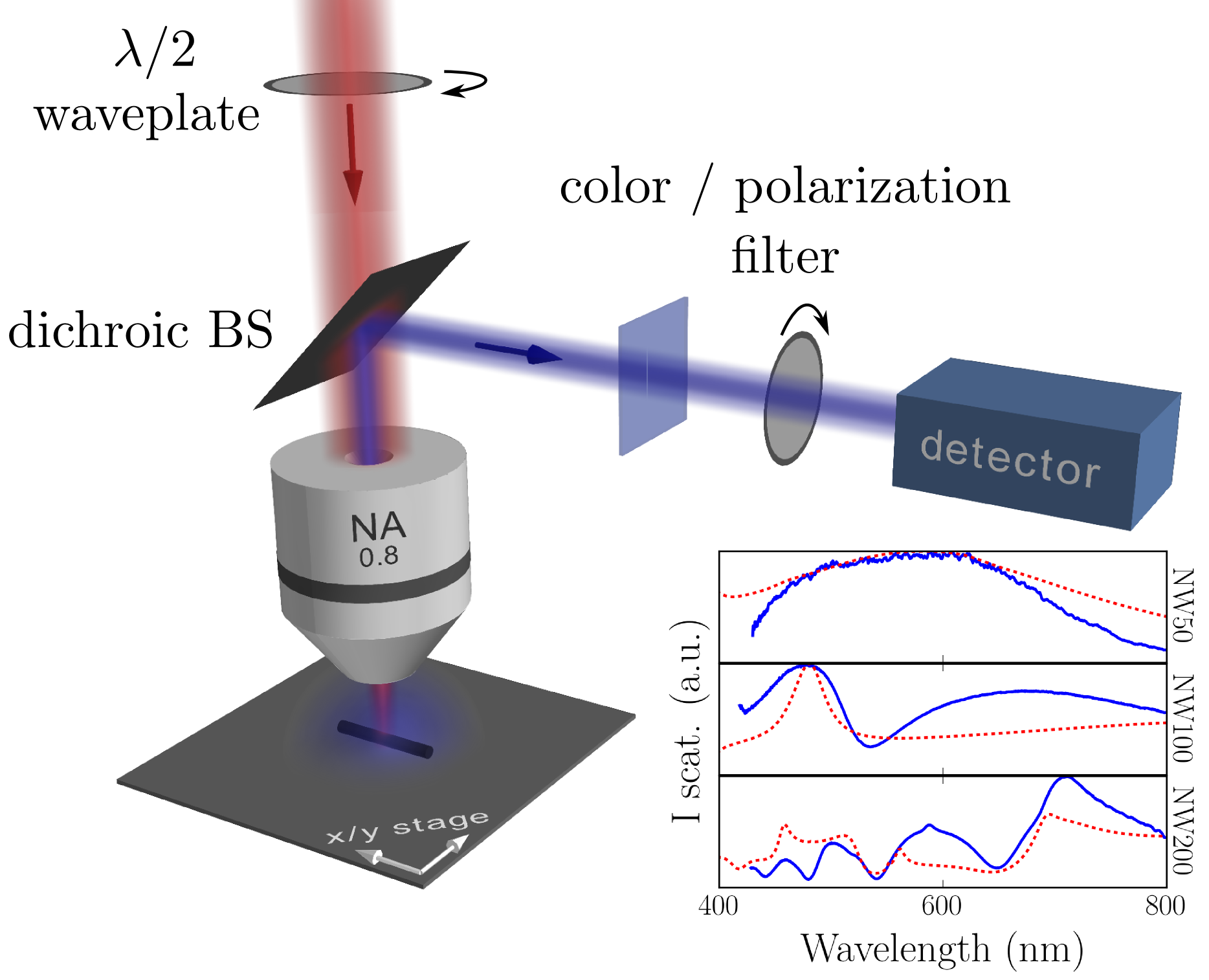} 
\caption{(color online) Scheme of the set-up for nonlinear experiments. The detection is composed of a photomultiplier tube in photocurrent mode connected to a lock-in amplifier. The inset shows typical linear scattering spectra for the three examined groups of nanowires and corresponding Mie scattering efficiencies.}
\label{fig:setup}
\end{figure}

$(111)$-oriented Si-NWs of different diameters were grown by the Vapor-Liquid-Solid method (VLS),\cite{cui_diameter-controlled_2001} and dispersed onto a fused silica substrate with lithographically-defined marks, allowing different measurements on a given NW.  The diameter $D$ of each NW was characterized by comparing its scattering spectrum to Mie theory.\cite{cao_tuning_2010,kallel_tunable_2012} 
This comparison, as well as neglecting longitudinal modes, is reliable for Si-NWs with length $L \gtrsim 3-4\,$\textmu m, significantly longer than their diameter.\cite{ee_shape-dependent_2015,traviss_antenna_2015}
In this work, we focus on three different groups defined by the diameter range: NW50 with diameters of about $50$\,nm, NW100 with diameters of about  $120$\,nm, and NW200  with diameters above $200$\,nm. Representative elastic light scattering spectra and corresponding Mie spectra for the three groups are shown in the inset of Fig.~\ref{fig:setup}.
NW50 shows one large non-degenerate TM$_{01}$ resonant mode in the visible (and thus no TE resonance), whereas NW100 displays two TM resonant modes and one TE mode (degenerate with the shorter wavelength TM mode).
Finally, the large NW200 have multiple TE/TM resonances.

A sketch of the experimental setup for SHG is given in Fig.~\ref{fig:setup}. It consisted of a Ti:Sa femtosecond (fs) laser at $\lambda=810\,$nm with a pulsewidth of about $150$\,fs and $80$\,MHz repetition rate, focused on the sample using a numerical aperture (NA) of $0.8$.
The average power at the back of the focusing objective was in the order of $10$\,mW.
Using a half-wave plate the linear polarization of the incident light could be aligned either parallel or perpendicularly to the NW axis, corresponding to TM or TE excitation, respectively. 
The sample was positioned with nm-precision using a piezo stage. 
The nonlinear emission was collected in epi-collection and reflected by a dichroic beam splitter to the detection system which consisted of a photomultiplier tube in photocurrent mode connected to a lock-in amplifier in order to increase the signal-to-noise ratio. The lock-in was synchronized with a mechanical chopper modulating the laser beam at 6 kHz.
Narrow-band color filters were introduced prior to the detector to select the Second Harmonic at $\lambda/2=405\,$nm, as well as a linear polarizer to analyze its polarization.

\section{Experimental Results}
\label{results}

\begin{figure*}[tb]
\centering
\includegraphics[width=\textwidth]{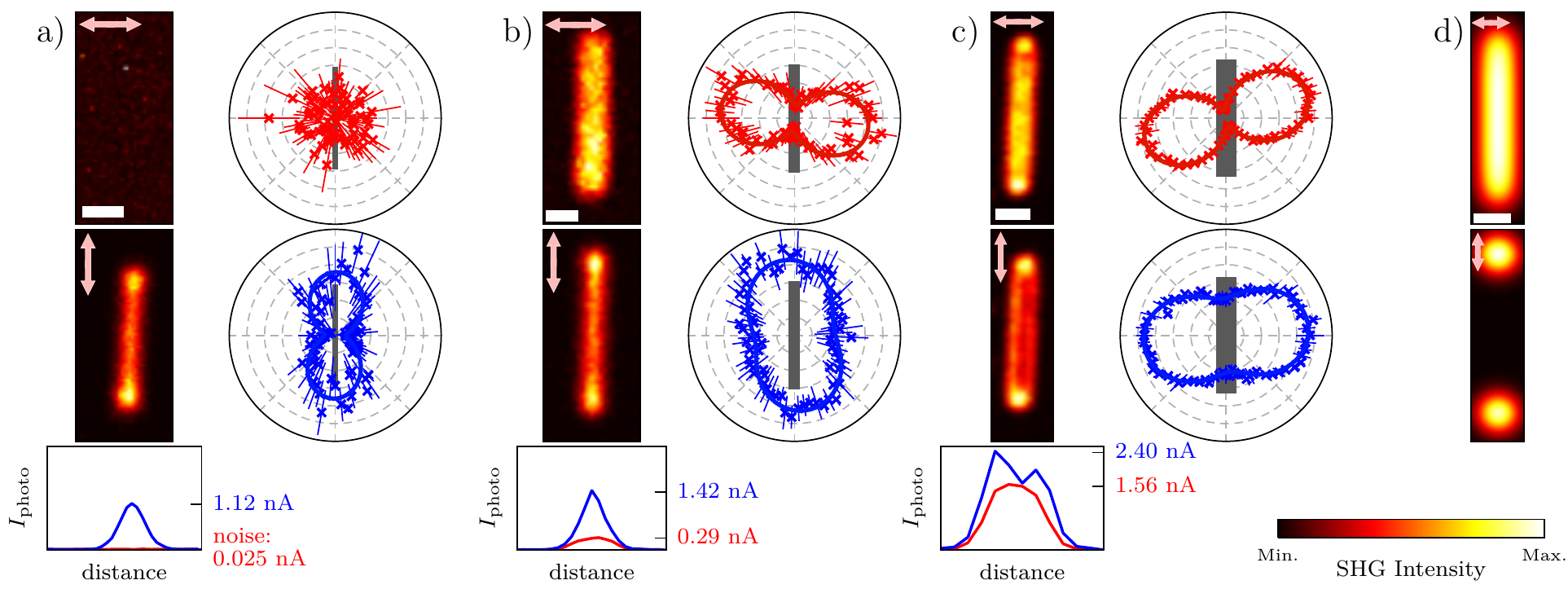}
\caption{(color online) Left columns: experimental SH maps (no polarization filter), each normalized to the maximum photocurrent intensity for a) NW50, b) NW100 and c) NW200. Right columns: SH polarization measurements (excitation at the NW center; solid lines are fits to the data). Bottom row: photocurrent profile measured across the NW center for TE (red) and TM (blue) configurations. d) Step-scan simulation for a $D=120\,$nm NW using $\mathbf{P}^{(2)}_{\perp \perp \perp}$ and TE (top) or TM (bottom) excitation. Scalebars are 0.5\,\textmu m. }
\label{fig:SH maps}
\end{figure*}

Figure~\ref{fig:SH maps} shows typical experimental results for representative NWs of a) NW50, b) NW100 and c) NW200 groups.
On the left of each subplot, the step-scan SH maps of the NWs are shown, whereas on the right SH polarization measurements recorded by irradiating the NW center are plotted. The laser spot size is about 620 nm for a wavelength of 810 nm.
As previously reported,\cite{wiecha_enhanced_2015} TE excitation produced homogeneous SHG along the NW, and TM excitation led to enhanced SHG from the tips. This trend is valid for all NW sizes except for the NW50 group, where in absence of a TE resonant mode, no SH was generated in the TE case.

\begin{figure}[tb]
\includegraphics[width=0.8\columnwidth]{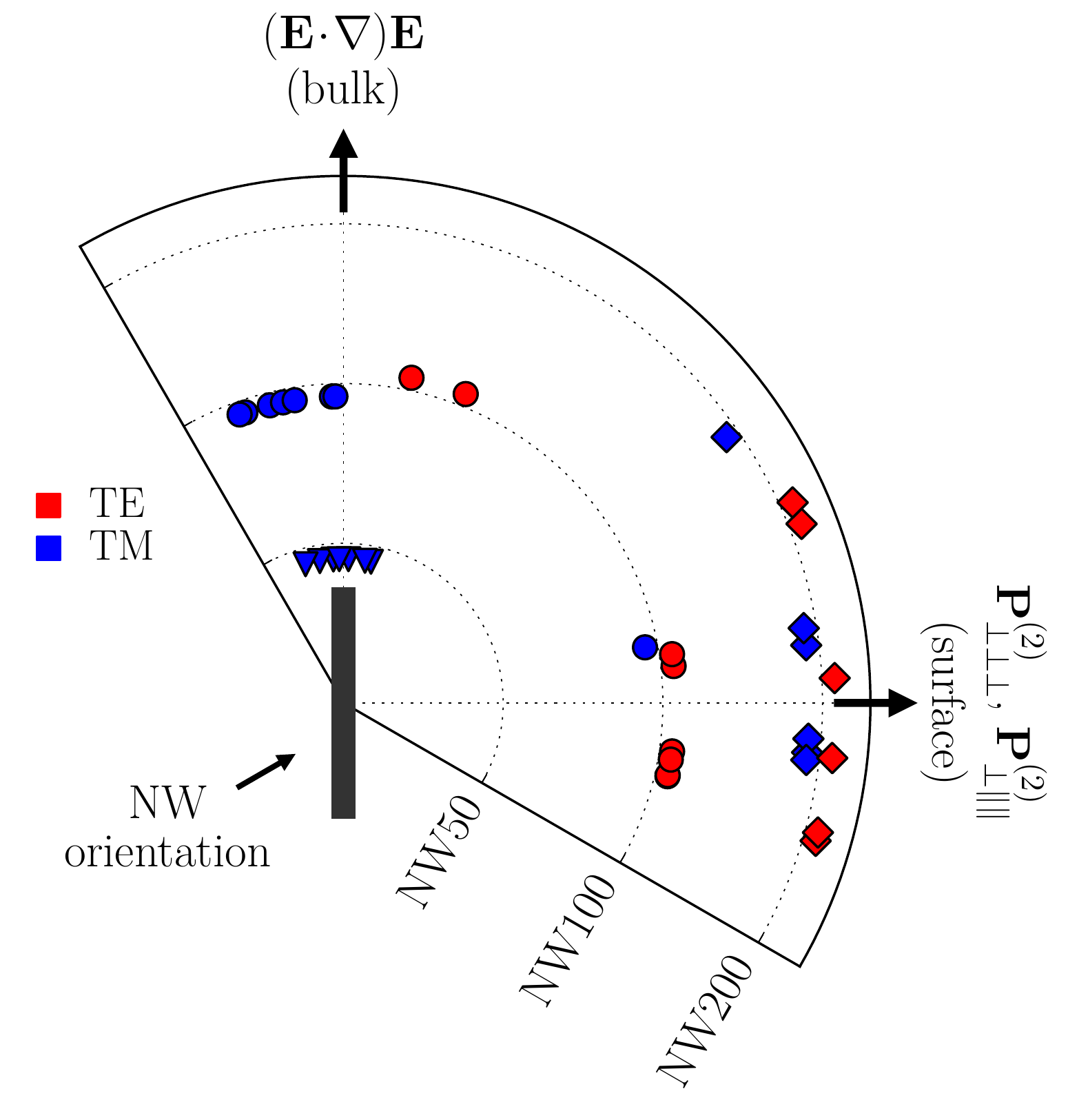}

\caption{(color online) Fitted SH polarization angles of NW50 (triangles), NW100 (circles) and NW200 (diamonds) for TE (red) and TM (blue) excitations focused at the NW center. The symbols are grouped per radial coordinate. The arrows indicate the direction of the far field SH polarization of the nonvanishing contributions. }
\label{fig:SH polarization}
\end{figure}

As shown in the SH polarization polar plots of Fig.~\ref{fig:SH maps}a-c, a 90$^{\circ}$ flip of the polarization direction was observed in the TM configuration. Contrary, under TE excitation, the SH light always followed the incident polarization, perpendicular to the NW axis. 
The general trend of SH polarization was confirmed by investigating over 20 different Si-NWs, which is shown in Fig.~\ref{fig:SH polarization}. 
Only a few NWs of the NW100 group showed atypical polarization behavior, which we attribute to possible partial illumination of one of the NW tips due to their relatively shorter length $L \approx 2\,$\textmu m. 
We observe nearly perfect figure-of-eight polar patterns in a few cases.  Qualitatively, a perfect figure-of-eight polar pattern is related to a linear SH polarization and a more or less open pattern implies that either the SH is elliptically polarized or the SH signal in the detector plane is inhomogeneous and has mixed polarization states. We will demonstrate that SHG from NW50 is due to a single contribution of the nonlinear susceptibility tensor and the nonlinear polarization rotation under TM excitation can be linked to two different contributions of this same tensor.

\section{Theoretical Considerations}
\label{theory}

In the following, we interpret our experiments by carrying out theoretical considerations on SHG from cylinders of centrosymmetric crystals. 
Second order electric polarization in centrosymmetric materials can be written as a superposition of both surface and bulk polarizations
\begin{equation}
 \mathbf{P}^{(2)}_{\text{cs}} = \mathbf{P}^{(2)}_{\text{sf}} + \mathbf{P}^{(2)}_{\text{bulk}}
\end{equation}

$\mathbf{P}^{(2)}$ labels the nonlinear polarization at the harmonic frequency $2\omega$. 
For isotropic, homogeneous media, $\mathbf{P}^{(2)}_{\text{sf}}$ can be expanded as three non-zero contributions\cite{heinz_second_1991}

\begin{subequations}\label{eq:Psurface}
 \begin{align}
   \mathbf{P}^{(2)}_{\perp \perp \perp} =         & \ \chi^{(2)}_{\perp \perp \perp} \left[ E_{\perp}^2 \right] \widehat{\mathbf{e}}_{\perp} \label{eq:SFnnn}\\
   \mathbf{P}^{(2)}_{\perp \parallel \parallel} = & \ \chi^{(2)}_{\perp \parallel \parallel} \left[ E_{\parallel}^2 \right] \widehat{\mathbf{e}}_{\perp} \label{eq:SFnpp}\\
   \mathbf{P}^{(2)}_{\parallel \parallel \perp} = & \ \chi^{(2)}_{\parallel \parallel \perp} \left[ E_{\perp} E_{\parallel} \right] \widehat{\mathbf{e}}_{\parallel} \label{eq:SFppn}
 \end{align}
\end{subequations}

$E$ represents the field amplitude at the fundamental frequency $\omega$, $\parallel$ and $\perp$ denote the directions parallel and perpendicular to the NW surface and for simplicity we set $\epsilon_0 = 1$.
Let us consider the case of an infinite cylinder. 
For an incident field normal to the cylinder axis (TE case), it turns out that the three surface terms lead to a nonlinear polarization perpendicular to the nanowire axis. 
This is obvious for equations~\eqref{eq:SFnnn} and \eqref{eq:SFnpp}. In addition, as $\widehat{\mathbf{e}}_{\parallel}$ in Eq.~\eqref{eq:SFppn} corresponds to $\widehat{\mathbf{e}}_{\varphi}$ in the cylindrical coordinate system, it is also perpendicular to the NW axis.
If the incident field is parallel to the axis (TM case), no field component $E_{\perp}$ normal to the cylinder surface exists, so that both $\mathbf{P}^{(2)}_{\perp \perp \perp}$ and $\mathbf{P}^{(2)}_{\parallel \parallel \perp}$ vanish. 
Thus equation~\eqref{eq:SFnpp} alone describes the surface SHG in the TM case, which is polarized along $\widehat{\mathbf{e}}_{\perp}$.
This leads to the insight, that under excitation far from the NW tips, surface SH polarization under either TE or TM excitation should always be perpendicular to the NW axis -- a finding that is in contradiction with the TM polar plots shown in Fig.\ref{fig:SH maps}a and Fig.\ref{fig:SH maps}b, where both SH and fundamental light polarizations are parallel to the NW axis.  
 
Let us therefore inspect the nonlinear bulk polarization, which -- as dipole-electric polarization is forbidden due to the lattice symmetry -- arise from field gradients in the material. Indeed, due to both the presence of leaky mode resonances (LMR) and a tightly focused laser beam, we presume that strong field gradients may be generated in the Si-NWs, so that bulk effects cannot be neglected. In first non-vanishing order, the bulk polarization consists of three terms\cite{heinz_second_1991}
\begin{equation}\label{eq:Pbulk}
 \mathbf{P}^{(2)}_{\text{bulk}} = 
      \gamma\, \nabla \left[\mathbf{E}^2\right] + 
      \beta\, \mathbf{E}\left[\nabla\cdot\mathbf{E}\right] + 
      \delta\, \left[\mathbf{E}\cdot\nabla\right]\mathbf{E}
\end{equation}
It has been shown that the $\gamma$-term can be included in equations~\eqref{eq:SFnnn} and \eqref{eq:SFnpp} using effective susceptibilities 
$\chi^{\text{eff.}}_{\perp\perp\perp} = \chi^{(2)}_{\perp\perp\perp} - \nicefrac{\gamma}{\left(\epsilon(\omega)\epsilon(2\omega)\right)}$ 
and 
$\chi^{\text{eff.}}_{\perp \parallel \parallel} = \chi^{(2)}_{\perp \parallel \parallel} - \nicefrac{\gamma}{\epsilon(2\omega)}$.\cite{guyot-sionnest_bulk_1988,dadap_optical_2008} 
Thanks to its surface-like behavior, it is often referred to as the non-separable bulk contribution, which becomes small for high-index semiconductors. 
We can also neglect the $\beta$-term in Eq.~\eqref{eq:Pbulk}, as $\nabla\cdot\mathbf{E}$ vanishes in the bulk of a homogeneous medium.\cite{cattaneo_polarization-based_2005,wang_surface_2009} Note that we omitted a term proportional to $E_i \nabla_i E_i$ whose susceptibility equals zero for homogeneous media.\cite{huo_second_2011}

Concerning the $\delta$-term in Eq.~\eqref{eq:Pbulk}, we find that under TE polarization strong field gradients appear only for large diameters because 
$(i)$ no field component exists along the axis, and 
$(ii)$ the in-plane fields normal to the axis can be considered constant for diameters below the appearance of the first resonant mode (at $\lambda=810\,$nm this is valid for $D\lesssim 150$\,nm, as shown in Appendix A).\bibnote[labelnote]{See Appendix: A) Electric field distribution in SiNWs by Mie theory ; B) Maps of the nonlinear electric field intensity distribution in the far field ; C) Surface SHG from Mie theory vs GDM simulations ; D) Cancellation of SH radiation from opposite dipoles in the far-field}
 In consequence, the last term in Eq.~\eqref{eq:Pbulk} is supposed to vanish for sufficiently small NWs in the TE configuration. 
Under TM illumination, field components normal to the cylinder axis are zero and the bulk polarization reduces to
\begin{equation}
  \mathbf{P}^{(2)}_{\text{bulk}, \text{TM}} = \delta \left( E_{z}\, \frac{\partial E_{z}}{\partial z} \right) \, \widehat{\mathbf{e}}_{z}
  \label{eq:PbulkTM}
\end{equation}
where $z$ denotes the axial direction. 
This means that for small NWs, the $\delta$-bulk contribution is the sole SH source able to generate a nonlinear polarization along the NW axis.
We consequently attribute SHG under TM excitation on the center of NW50 and NW100 (figures~\ref{fig:SH maps}a, \ref{fig:SH maps}b and \ref{fig:SH polarization}) mainly to the $(\mathbf{E}\cdot\nabla)\mathbf{E}$ bulk contribution. For larger NWs the surface term becomes more significant, leading to the observed flip of the polarization (figures~\ref{fig:SH maps}c and \ref{fig:SH polarization}.

In summary, three contributions to SHG remain under consideration to explain our experimental results: the surface $\perp\perp\perp$ and $\perp\parallel\parallel$ terms, resulting in a nonlinear polarization perpendicular to the NW axis, and the $(\mathbf{E}\cdot\nabla)\mathbf{E}$ bulk source, creating a polarization along the NW axis. 
$\chi^{(2)}_{\perp\perp\perp}$ is usually considered to dominate SHG, while the weaker surface terms ($\chi^{(2)}_{\perp\parallel\parallel}$, $\chi^{(2)}_{\parallel\parallel\perp}$) and the separable bulk susceptibilities are supposed to be of comparable magnitude.\cite{falasconi_bulk_2001} We have now to examine the dependence of these different contributions on the NW diameter. This will be done in the following using electrodynamical simulations with the Green Dyadic Method (GDM).

\section{Simulations of Second Harmonic Generation using the Green Dyadic Method}
\label{Sim}

\begin{figure}[tb]
\centering
\includegraphics[width=\columnwidth]{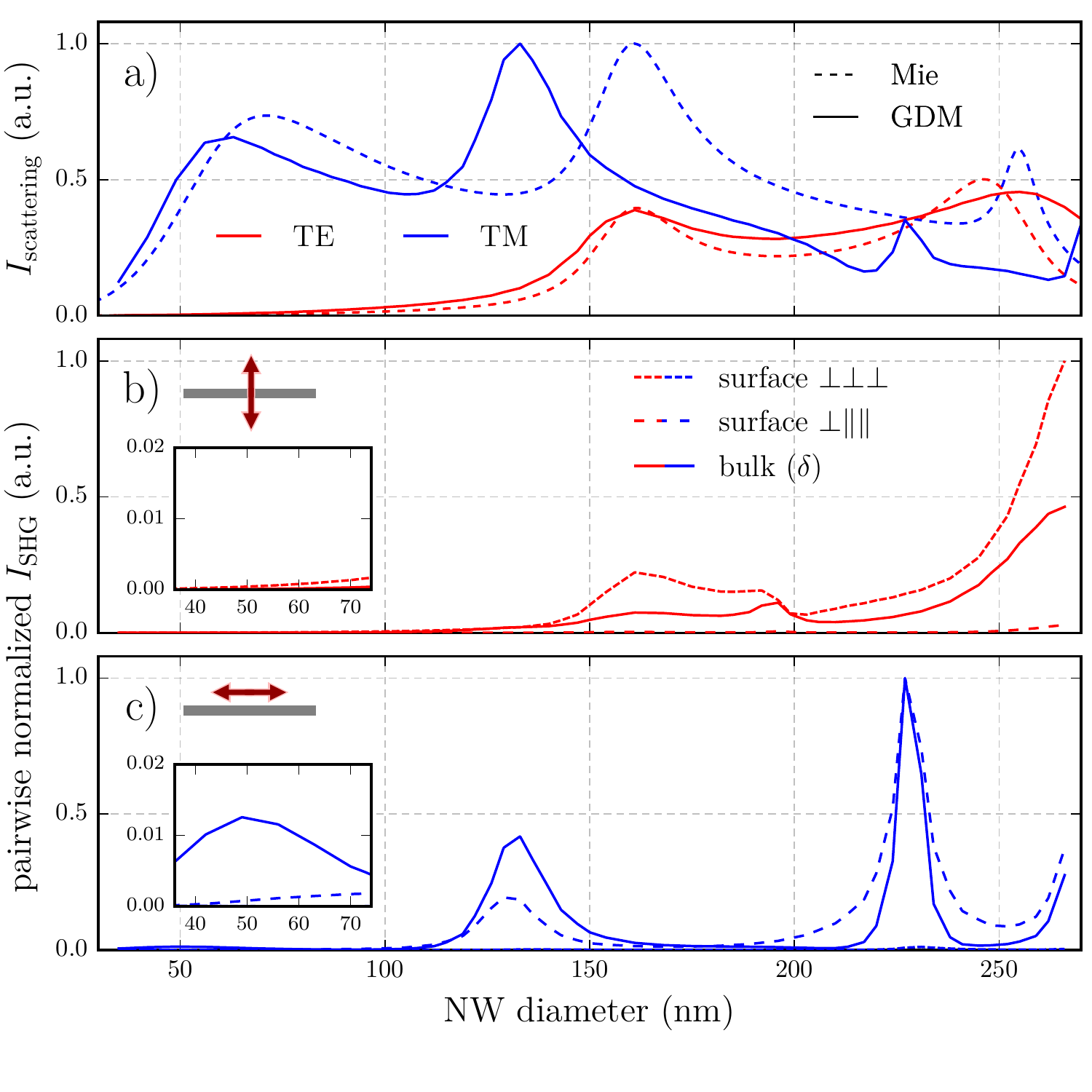}
\includegraphics[width=\columnwidth]{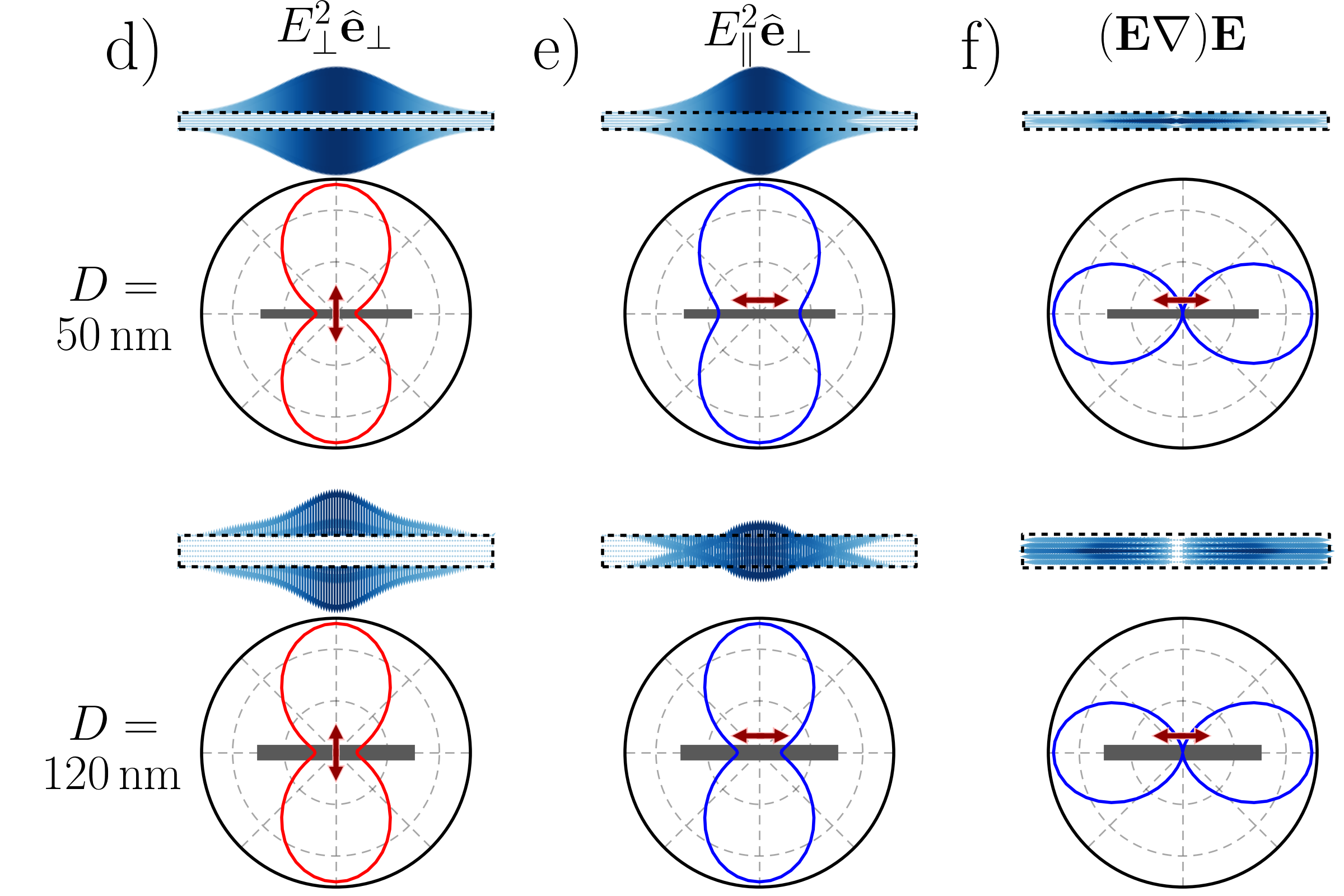}
\caption{(color online) All data for an incident wavelength of $\lambda=810$\,nm. (a) Elastic scattering intensities from Mie theory (dashed) and GDM simulations (solid) for TE (red) and TM (blue) excitation. 
GDM simulated SHG far-field intensities are plotted in (b) for TE and (c) for TM excitation. Surface (pointed: $\perp\perp\perp$, dashed: $\perp\parallel\parallel$) and bulk SHG ($\delta$-term, solid lines) are pairwise normalized to their overall (TE / TM) maximum. The insets show zooms on the region of small diameters.
d-f show the nonlinear polarizations $\mathbf{P}^{(2)}$ (real parts, dense vector plots in blue color) and SH far-field polarizations (polar plots) for (d) TE excited $\perp\perp\perp$, (e) TM excited $\perp\parallel\parallel$ and (f) TM excited bulk ($\delta$) for a $D=50$\,nm and a $D=120$\,nm NW (top and bottom respectively).}
\label{fig:simulations}
\end{figure}

\subsection{Simulation Method and Model} 
The GDM applies the principle of a generalized propagator to calculate the electric field in an arbitrarily shaped nanostructure.\cite{martin_generalized_1995}
This approach, succesfully applied in nano-plasmonics, is particularly suitable for step-scan simulations, as the generalized propagator is independent of the incident electric field and hence has to be calculated only once for each wavelength.\cite{teulle_scanning_2012} 
The incident field at frequency $\omega$, modeled as a focused beam (NA\,0.8), is then numerically raster-scanned over the model. 
A further advantage of the GDM is that the presence of a substrate (in our case $n_{\text{subst.}}=1.5$) can be taken into account at no additional computational cost.
Subsequently, the fundamental field inside the structure is used to calculate $\mathbf{P}^{(2)}$ on the surface and in the bulk using Eq.~\eqref{eq:Psurface} and Eq. \eqref{eq:Pbulk}. Central differences are used to approximate the gradients in Eq.~\eqref{eq:Pbulk}.
Finally, each meshpoint is considered as an emitter at $2\omega$ and the radiation of the ensemble to the far-field is calculated using a propagator taking into account the presence of the substrate.\cite{novotny_allowed_1997} The far-field intensity is integrated over the collecting solid angle (NA\,0.8) and optionally analyzed for its polarization. In Appendix B are shown the maps of the electric field intensity distribution in the far field. \bibnote[labelnote]{See Appendix B}
The fundamental wavelength was set to $\lambda=810$\,nm, consequently the harmonic radiation was calculated at $\lambda/2=405$\,nm.

In order to simplify the numerical work, simulations have been performed using wires of rectangular cross section (See Appendix C).\bibnote[labelnote]{See Appendix B}  It has been shown, that the difference between cylindrical and rectangular sections only shifts the resonances, while the resonance number is conserved.\cite{van_de_groep_designing_2013}
The validity of this assumption is verified when comparing Mie theory for an infinite cylinder to simulated elastic scattering spectra of our model (Fig.~\ref{fig:simulations}a). To allow comparison with Mie theory, the simulated wires were chosen to be long compared to the spot size of the incident beam ($L>2$\,\textmu m).

\subsection{$\mathbf{P}^{(2)}_{\perp\perp\perp}$ Surface contribution}
In Fig.~\ref{fig:SH maps}d, an image of a SHG raster-scan simulation, considering the  $\mathbf{P}^{(2)}_{\perp\perp\perp}$ surface term only, is shown for a NW of $D=120$\,nm. 
Obviously, the global trend of homogeneous SHG for TE and tip-enhanced SHG for TM can be reproduced using only this normal surface contribution. Similar results are obtained for smaller and larger diameters.
However, as previously pointed out, two experimental phenomena can not be explained by only Eq.~\eqref{eq:SFnnn}, which are TM excited SHG from the NW center and SH polarized along the NW axis.

\subsection{Diameter-dependence of SHG contributions}
In order to verify the hypothesis of mainly $\mathbf{P}^{(2)}_{\perp\perp\perp}$ generated-SH in the TE case on the one hand and mixed $\perp\parallel\parallel$-surface / $\delta$-bulk SH for the TM case on the other hand, diameter-dependent SHG simulations, shown in figure~\ref{fig:simulations}b-f, were carried out. 
A focused (NA\,0.8) incident electric field at $\lambda = 810$\,nm, either polarized TE (Fig.~\ref{fig:simulations}b) or TM (Fig.~\ref{fig:simulations}c), was set on the center of a NW model, whose section was progressively increased. 
SH intensities in the far-field were calculated for the $\mathbf{P}^{(2)}_{\perp\perp\perp}$ and $\mathbf{P}^{(2)}_{\perp\parallel\parallel}$ surface terms, as well as for the $\delta$-bulk contribution. Each contribution was normalized to the highest intensity value within both incident polarizations. This means, that absolute comparison of SH intensities is only possible for each contribution, separately.

The $\perp\perp\perp$-surface contribution under TE excitation exceeds the case of TM incidence on the entire diameter range by several orders of magnitude. 
As $\chi^{(2)}_{\perp\perp\perp}$ is supposed to surpass the other second order susceptibility components, we conclude that SHG under TE excitation is dominated by the normal surface component, whereas under TM excitation, the $\perp\perp\perp$-surface contribution seems to be negligible over the whole simulated range, which is in agreement with the theoretical prediction.

While the normal surface term vanishes for incident fields along the axis, $\perp\parallel\parallel$-surface and $\delta$-bulk contributions are radiated more efficiently than in the TE case. 
We also see in Fig.~\ref{fig:simulations}c that the surface term grows more rapidly with increasing diameters when compared relatively to the bulk term.
This supports our assumption that SHG from TM illumination on the NW center is due to $\mathbf{P}^{(2)}_{\perp\parallel\parallel}$ and/or $\delta$-bulk contributions, depending on the diameter range. 
We show in Fig.~\ref{fig:simulations}d-f simulated $\mathbf{P}^{(2)}$ near-fields (top row) and their polarization patterns after radiation to the far-field (bottom row) for $D=50$\,nm and $D=120$\,nm. The behavior of the SH polarization is identical for all sizes of simulated wires.
The $\mathbf{P}^{(2)}_{\perp\perp\perp}$ case under TE excitation shown in  Fig.~\ref{fig:simulations}d is in agreement with the experimental results. $\mathbf{P}^{(2)}_{\perp\parallel\parallel}$ and the $\delta$-bulk term under TM excitation are shown in Fig.~\ref{fig:simulations}e and Fig.~\ref{fig:simulations}f, respectively. These simulations show the rotation of of the far-field polarization pattern with respect to the NW axis. This is in agreement with the experimental plots of Fig.~\ref{fig:SH maps}b-c and confirms ultimately the axis-parallel polarization emitted by the $\delta$-bulk term, which is also the main contribution to SHG from NW50. 

It is rather counterintuitive that SHG in small diameter nanowires occurs due to the $\delta$-bulk contribution while the surface sources increase with increasing diameter -- hence for decreasing surface over volume ratio.
Resonant optical modes have an influence on the nonlinear contributions, as shown in figure~\ref{fig:simulations}a-c. In particular, the fundamental  mode TM$_{01}$  leads to a homogeneous field enhancement within the NW section, which has a direct consequence on the $\delta$-bulk term (see equation~\eqref{eq:PbulkTM}). 
However, the relation between  fundamental and harmonic local field enhancement is not a sufficient condition to explain the efficiency of the SH radiation to the far-field, as strong silencing is expected due to the high symmetry of the nanowires.\cite{neacsu_second-harmonic_2005,berthelot_silencing_2012} 
By analyzing the nonlinear polarization vectors (figures~\ref{fig:simulations}d-f), we indeed find a strong microscopic cancellation for the surface contributions while retardation among the bulk polarization vectors neutralizes the cancellation of oppositely radiating dipoles to the farfield (see also Appendix C).\bibnote[labelnote]{Appendix C}

\section{Conclusion}

In conclusion, our study of SHG from individual Si-NWs showed that $\mathbf{P}^{(2)}_{\perp \perp \perp}$ dominates SHG for TE polarized excitation, resulting in a SH polarization normal to the NW axis, which is in agreement with former observations of $\chi^{(2)}_{\perp\perp\perp}$ as leading source of second-order susceptibility. 
For TM excitation on the other hand, $\mathbf{P}^{(2)}_{\perp \perp \perp}$ vanishes as soon as the laser spot leaves the NW tips, giving the opportunity to examine the $\mathbf{P}^{(2)}_{\perp\parallel\parallel}$ surface source and the $\delta$-bulk contribution in more detail.
A diameter-dependent flip of the SH polarization was observed in this case, which we studied using numerical simulations. The latter confirmed a changeover in leading order contribution from bulk ($(\mathbf{E\nabla})\mathbf{E}$) SHG for  small NWs to surface ($\mathbf{P}^{(2)}_{\perp\parallel\parallel}$) SHG for larger NWs with diameters $\gtrsim 150$\,nm. 
We concluded that radiation from both $\mathbf{P}^{(2)}_{\perp\parallel\parallel}$ and $\mathbf{P}^{(2)}_{\delta,\text{bulk}}$ is of comparable magnitude and can be individually addressed by simply adjusting the diameter of the nanowire, which is particularly interesting as the $\delta$-bulk contribution is supposed to be difficult to isolate from the other SHG terms.

We showed that, because of their geometry and optical properties, Si-NWs provide a highly promising research platform to gain insight in the relations between surface and bulk contributions of SHG from centrosymmetric materials in general.
This might also allow to get estimate values for the different contributing $\chi^{(2)}$ terms, though accurate quantification of the different $\chi^{(2)}$ elements may be a rather difficult task, due to strong silencing of the nanoscopic nonlinear polarization because of the high symmetry of the NWs.\\

\begin{acknowledgments}
This work was supported under ``Campus Gaston Dupouy'' grant by French government, R\'egion Midi-Pyr\'en\'ees and European Union (ERDF), and by the computing facility center CALMIP of the University of Toulouse under grant P12167.
\end{acknowledgments}

\bibliography{bibliography}

\clearpage

\section*{APPENDIX}
%
%
%
%
\setcounter{equation}{0}
\setcounter{figure}{0}
\setcounter{table}{0}
\setcounter{page}{1}
\makeatletter
\renewcommand{\theequation}{A.\arabic{equation}}
\renewcommand{\thefigure}{A.\arabic{figure}}
\renewcommand{\bibnumfmt}[1]{[S#1]}
\renewcommand{\citenumfont}[1]{S#1}
%
%
%
\subsection{Electric field distribution in Si-NWs by Mie theory}
\begin{figure}[htb]
\centering
\includegraphics[width=1.0\columnwidth]{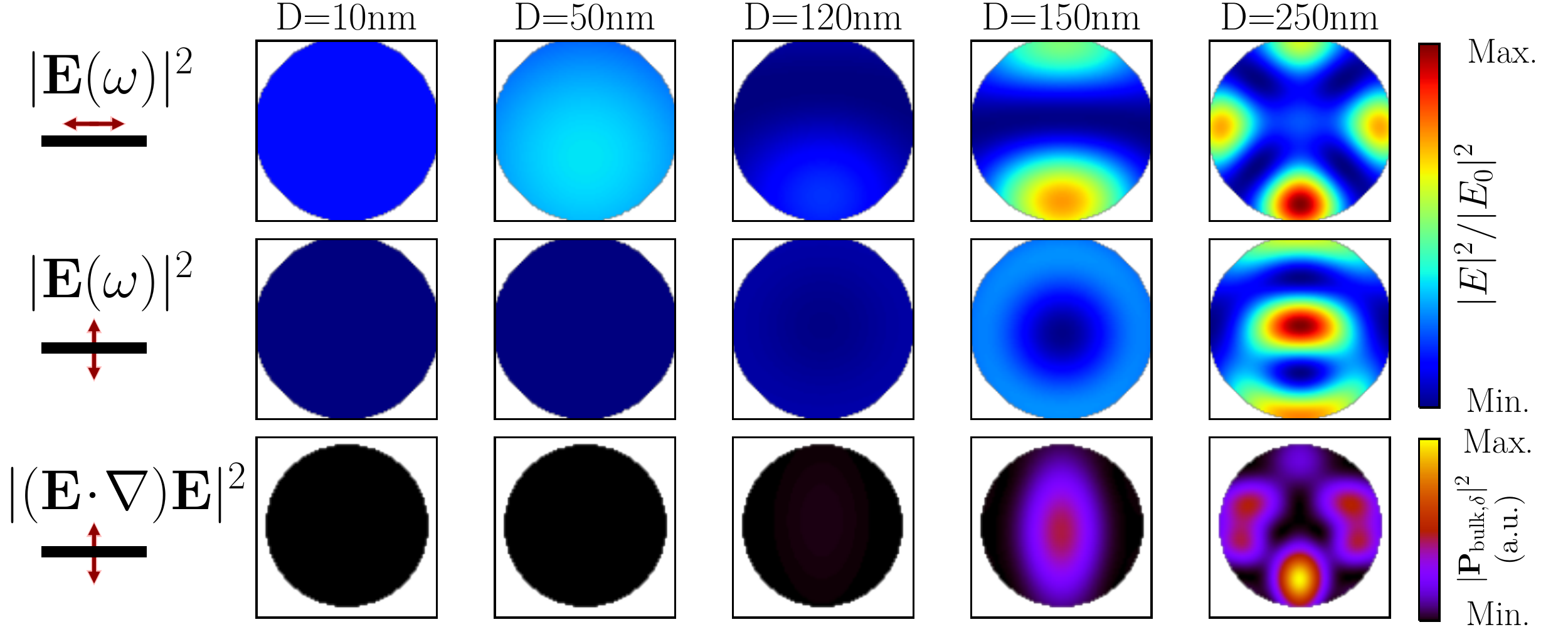} 
\caption{(color online) Fundamental electric field intensity distribution as function of Si-NW diameter for TM (top row) or TE  (center row) configuration. Non-linear field intensity distribution from the bulk-like polarization $(\mathbf{E \nabla})\mathbf{E}$ in the TE case (bottom row). Incident plane-wave from the top.}
\label{fig:SIfielddistribution}
\end{figure}
Field distributions shown in Fig.~\ref{fig:SIfielddistribution} were calculated using Mie theory for an infinite cylinder under plane-wave excitation at 810\,nm. 
The uniform and near zero field intensity in NWs having a diameter smaller than 150\,nm in the TE case confirms that there is no bulk contribution $(\mathbf{E \nabla})\mathbf{E}$ to SHG for NW50 and NW100. 
In such case, the non-linear emission under TE polarized excitation is due to surface effects only. For nanowires of larger diameter however, they can arise from both bulk and surface contributions.
 
Concerning the TM case in Mie theory, even though the field intensity is high for small nanowires (presence of the fundamental mode), no field gradient along the NW axis can exist under plane-wave illumination. 
The bulk contribution can therefore not be calculated for TM excitation and is thus not shown here. 
The field gradient introduced by a tightly focused excitation beam can only be modeled numerically.

\subsection{Maps of the nonlinear electric field intensity distribution in the far field}
\begin{figure}[htb]
\centering
\includegraphics[width=1.0\columnwidth]{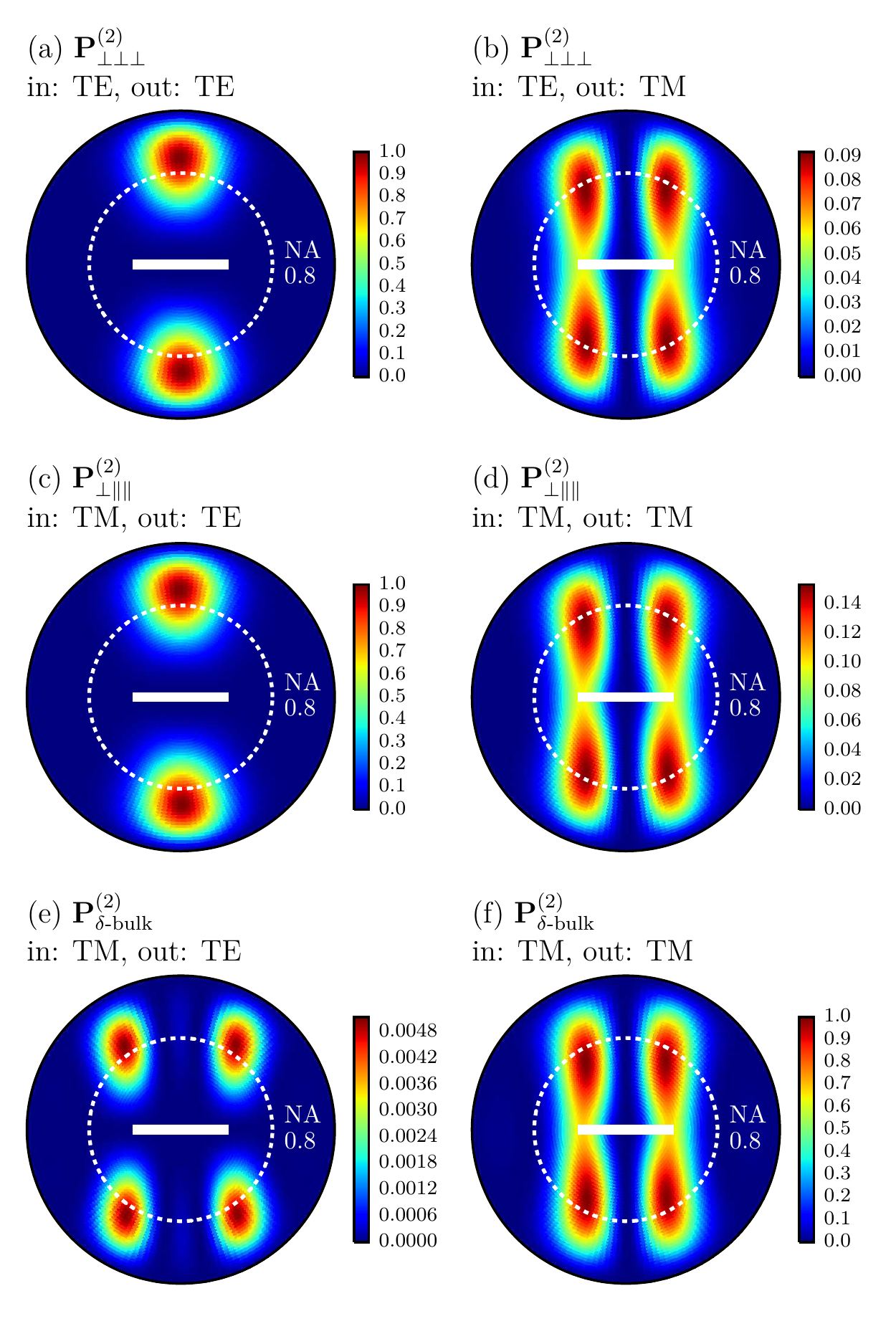} 

\caption{(color online) Maps of the nonlinear electric field intensity distribution in the far field (half-upper space) for NW50. 
The incident TE or TM polarization is given by ``in'', with respect to the NW axis represented by the white bar. 
The nonlinear electric field is analyzed in two directions (``out''): perpendicular to the axis (left column), and parallel to the axis (right column). 
The normalization is done using the maximum intensity within each case. The dashed line circle shows the detection limit due to the objective NA\,0.8.}
\label{fig:SISHfielddistribution}
\end{figure}

The maps shown in Fig.~\ref{fig:SISHfielddistribution} are qualitatively the same for the different NW diameters investigated in the article. 
The total intensity on the detector corresponds to the integrated intensity over the area delimited by the objective NA (dashed circle). 
As expected due the symmetry of the system, the intensity in the center of the map is zero.
Taking the figures of the lower row, the integrated intensity for TM$^{\text{in}}$-TM$^{\text{out}}$ is much higher than in the  TM$^{\text{in}}$-TE$^{\text{out}}$ case. 
This results in almost perfectly closed figure-of-eight patterns, as shown in Fig.~\ref{fig:SH polarization}a and Fig.~\ref{fig:simulations}f of the main article. 
The relatively high ratios between TE$^{\text{out}}$ and TM$^{\text{out}}$ for the case of surface SHG show, that the figure-of-eight patterns can be more or less open also for a single contribution, as shown in Fig.~\ref{fig:simulations}d-e.
The SHG intensity-distribution in the farfield also shows that reducing the objective NA could enhance the detection of the $\delta$-bulk contribution with respect to the surface components.

\vfill\eject
\subsection{Surface SHG from Mie theory vs. Green Dyadic Method simulations}

\begin{figure}[htb]
\centering
\includegraphics[width=1.0\columnwidth]{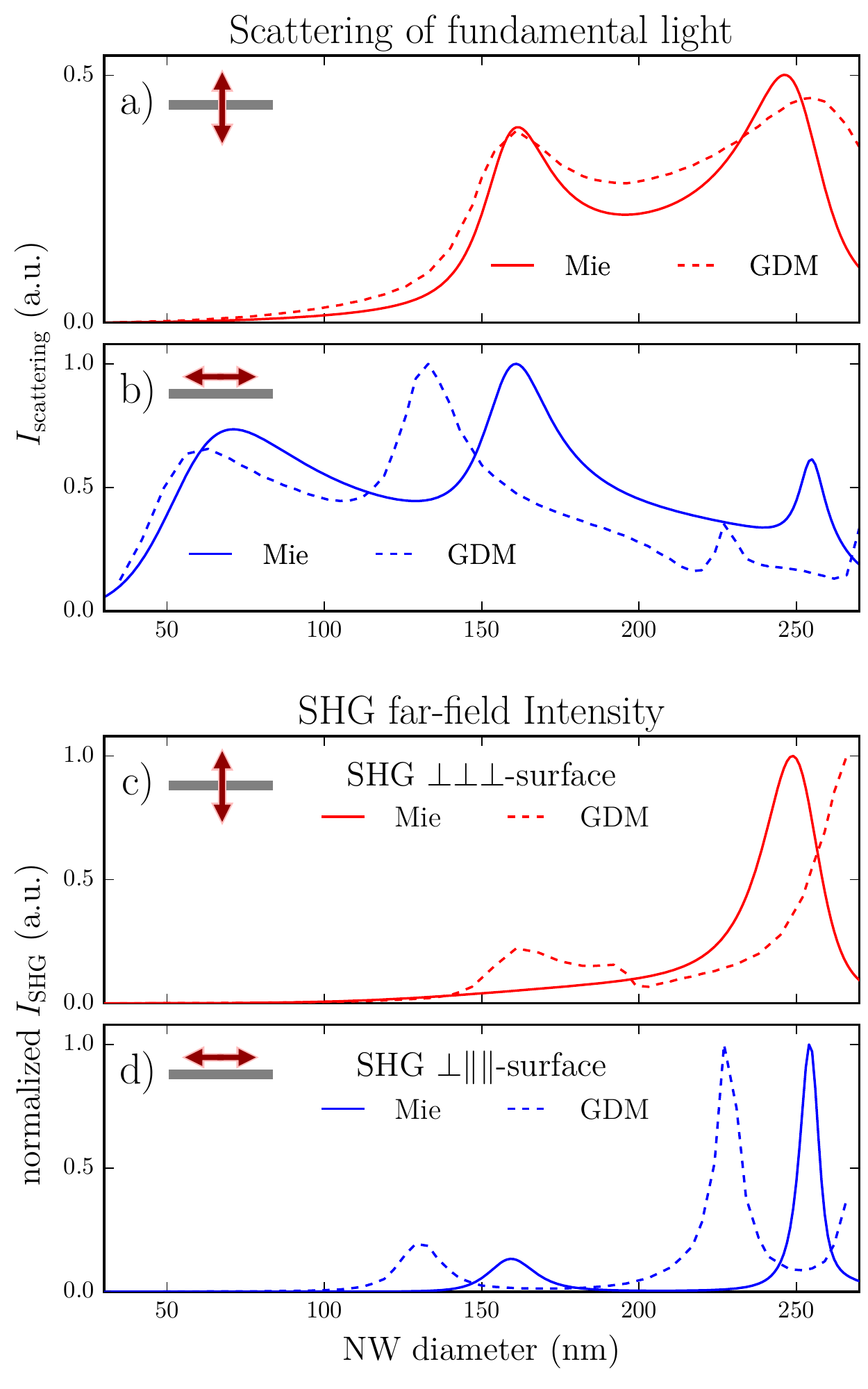} 

\caption{(color online) Scattering intensities (a, b) and surface SHG intensities (c, d) calculated analytically by Mie theory and numerically by GDM (solid and dashed lines, respectively). a,c): TE polarization and b,d): TM polarization.}
\label{fig:SImievsgdmshg}
\end{figure}
In figure~\ref{fig:SImievsgdmshg} linear and nonlinear scattering intensities are compared. They are calculated by Mie theory for plane-wave illuminated infinite cylinders on the one hand and using the Green Dyadic Method (GDM) for finite length wires of rectangular cross-section under tightly focused excitation on the other hand.
Apart from a shift of the resonances and a slight change of some resonance widths, both approaches yield similar results, showing a correlation between linear and nonlinear intensities. 
The comparison of both methods in the case of surface contributions to SHG validates the use of the numerical method to evaluate further nonlinear contributions that cannot be obtained using Mie theory.

\subsection{Cancellation of radiation from opposing dipoles in the far-field}

\begin{figure}[htb]
\centering
  \setbox1=\hbox{\includegraphics[height=2.8cm]{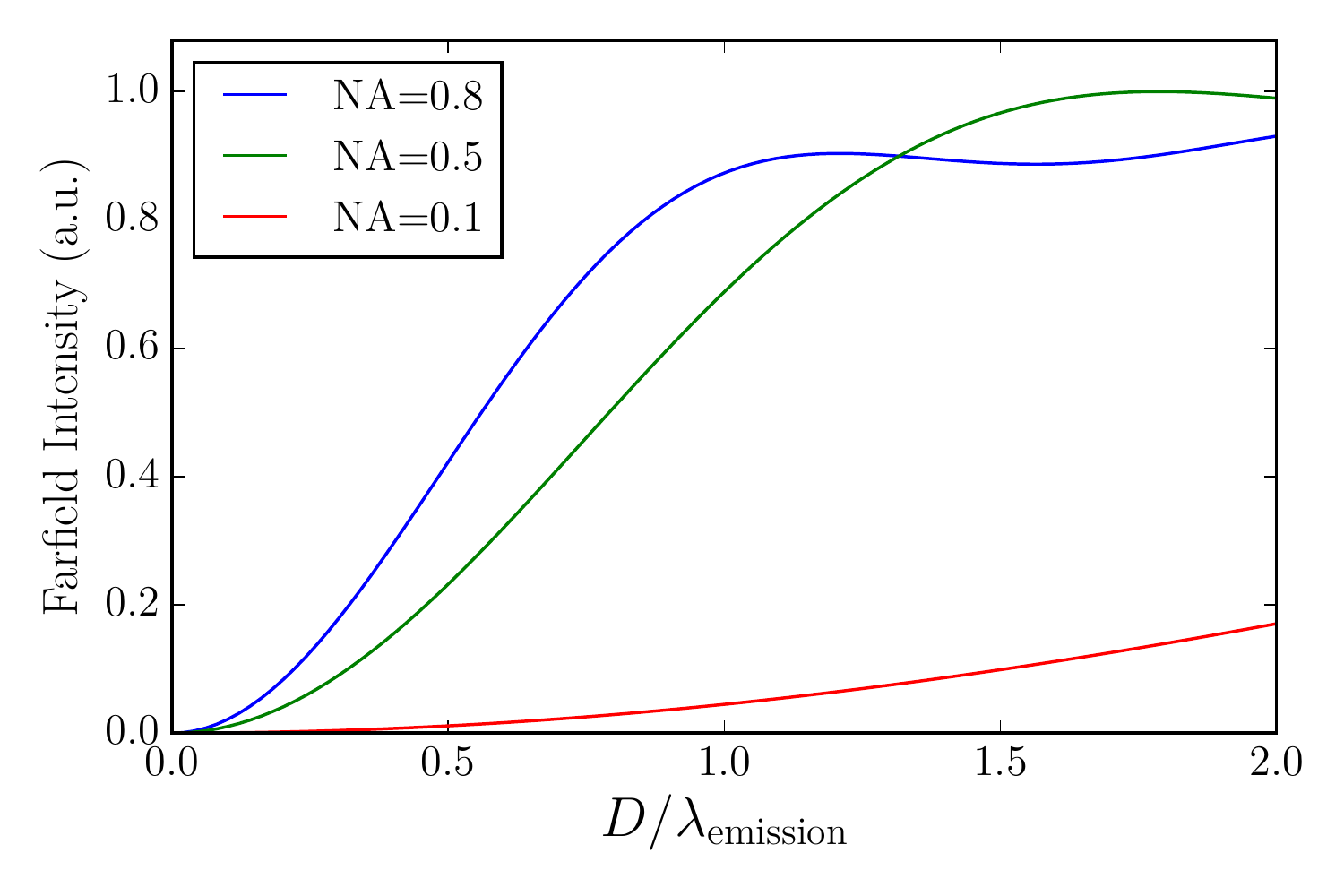}}
  \includegraphics[width=0.9\columnwidth]{figSI_04}\llap{\makebox[\wd1][c]{\includegraphics[width=.18\columnwidth]{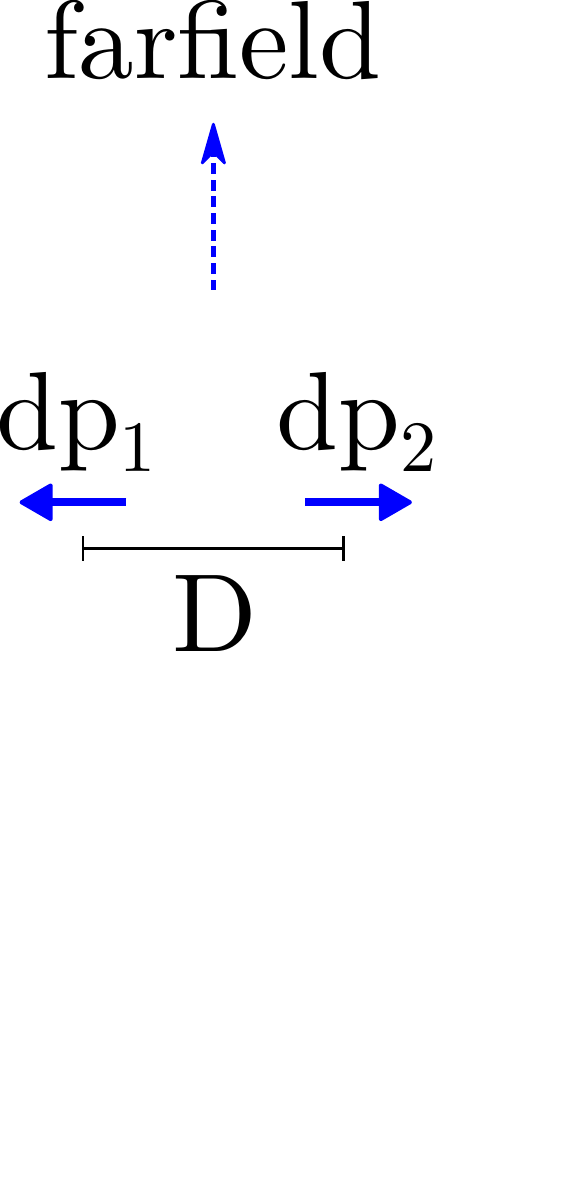}}} 
\caption{(color online) Far-field intensity of two coherently radiating dipoles of equal amplitude and opposite phase  in normal direction to their polarization vector as function of the distance between the two dipoles. The intensity is integrated over different solid angles where NA\,0.8 corresponds to the objective used in the experiments. The inset shows a sketch of the considered geometry.}
\label{fig:SIdipolesFF}
\end{figure}
To give an illustrative explanation for the somehow counterintuitive observation of bulk effects dominating for small nanowires while surface effects occur only for larger structures, we consider two oppositely oscillating dipoles, as found for the nonlinear surface polarizations $\mathbf{P}^{(2)}_{\perp\perp\perp}$ and $\mathbf{P}^{(2)}_{\perp\parallel\parallel}$. 
The far-field radiation intensity through solid angles corresponding to different numerical apertures is plotted as a function of the inter-dipole distance in figure~\ref{fig:SIdipolesFF}. While for small distances cancellation is almost perfect, the radiation becomes observable in the far-field only for distances corresponding to a major fraction of the wavelength. 

\begin{figure}[htb]
\centering
  \includegraphics[width=0.9\columnwidth]{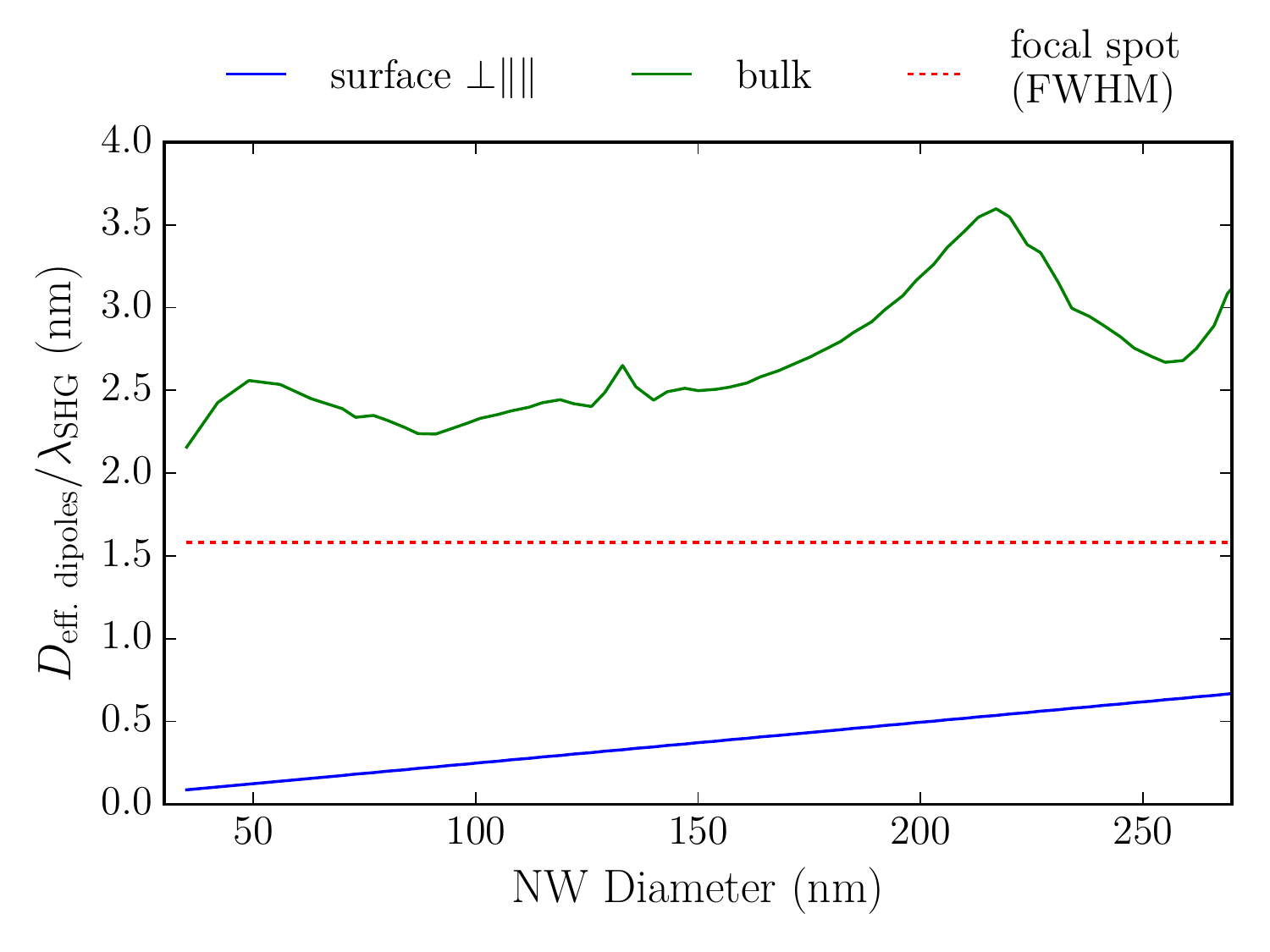}
\caption{(color online) Distance of two ``effective'' dipoles (see text) for surface (blue) and bulk (green) nonlinear polarization under TM excitation, calculated from the simulation data. The dips observed for the bulk dipole-distance correspond to the resonance positions (compare with fig.~\ref{fig:simulations} of the main text).}
\label{fig:SIdipolesFFeffDPTM}
\end{figure}

The bulk polarization is mainly induced by the field gradient from tight focusing. Hence the average distance of dipoles oscillating with opposite phase will mainly be determined by the focal spot size which is of constant value.
\begin{figure}[htb]
\centering
  \includegraphics[width=0.7\columnwidth]{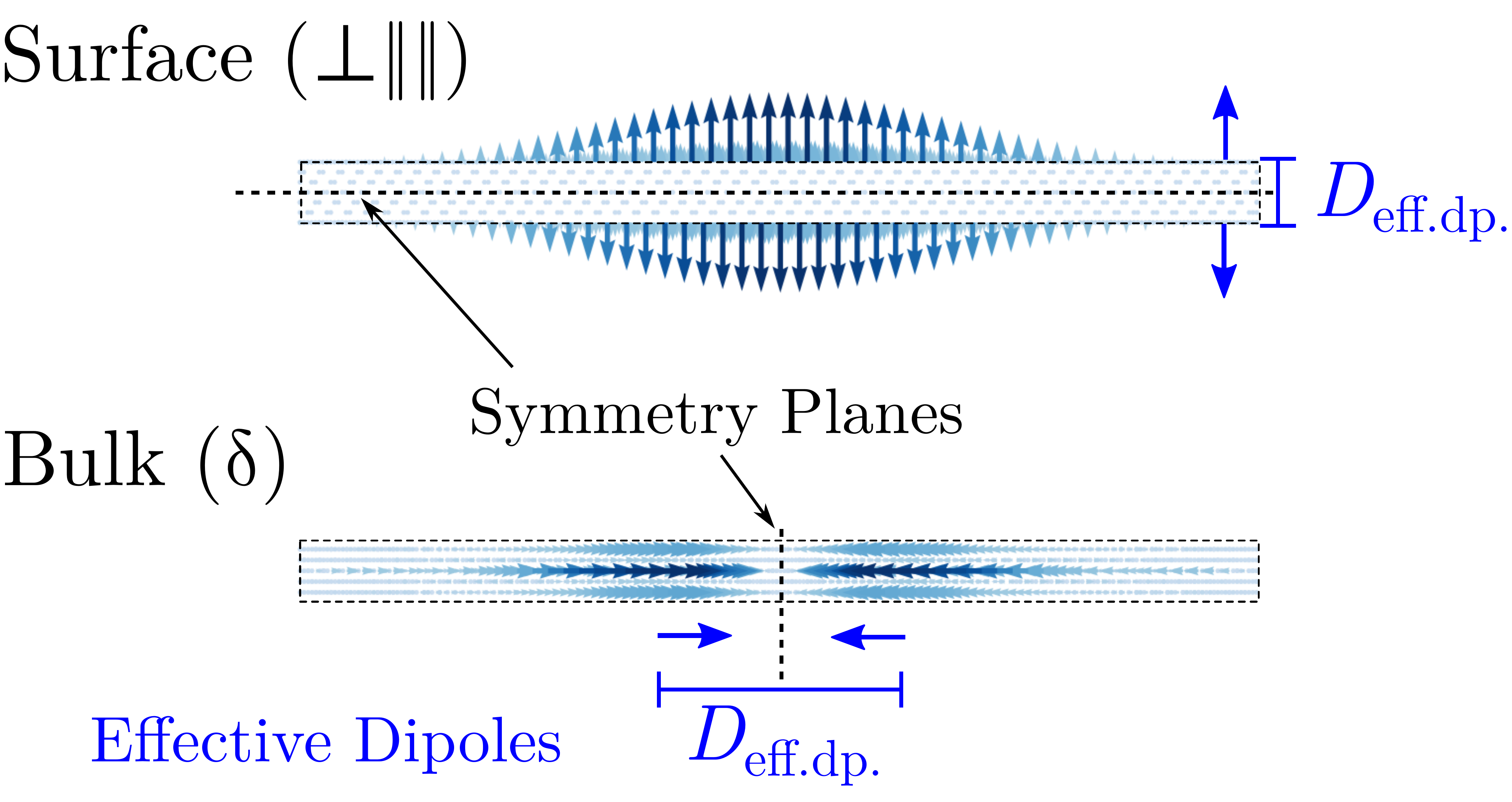}
\caption{(color online) Sketch of the data analysis. The nonlinear polarization is separated in two areas with respect to a symmetry plane. In each area, the average dipole is calculated, neglecting retardation effects. Radiation towards the reader.}
\label{fig:SIdipolesFFeffDPsketch}
\end{figure}
In order to verify that this assumption is valid for the case of $\mathbf{P}^{(2)}_{\perp\parallel\parallel}$ and $\mathbf{P}^{(2)}_{\text{bulk}, \delta}$ in TM excited nanowires, we reduce the nonlinear polarization to two effective dipoles, oscillating with opposite phase. We define their positions using the weighted averages 
\begin{equation}
 \mathbf{r}_{\text{eff.}} = \frac{\sum\limits_i \mathbf{r}_i |\mathbf{P}^{(2)}_i|}{\sum\limits_i |\mathbf{P}^{(2)}_i|}
\end{equation}
of all dipoles $\mathbf{P}^{(2)}_i$ at $\mathbf{r}_i$ in two symmetric fractions of the structure as illustrated in figure~\ref{fig:SIdipolesFFeffDPsketch}. In this rough approximation we neglect retardation effects in the summation by taking the modulus of each complex polarization vector. The distance between the two effective dipoles is plotted in figure~\ref{fig:SIdipolesFFeffDPTM} as a function of the nanowire diameter.
The surface polarization has always an effective spacing corresponding to the nanowire diameter. The $\delta$-bulk nonlinear polarization on the other hand is found to be characterized by two effective dipoles with a separating distance always larger than the focal spot size. This behavior is in agreement with our initial hypothesis and can explain the observation of surface effects for large NW diameters only, while bulk SHG is observed for small nanowires.


\clearpage

\end{document}